\documentclass{aa}  

\usepackage{graphicx}
\usepackage{txfonts}
\usepackage{booktabs,caption}
\usepackage{threeparttable}
\usepackage{graphicx}
\usepackage{multicol}
\begin{document}

   \title{Effect of finite disk-thickness on swing amplification of non-axisymmetric perturbations in a sheared galactic disk}
	\titlerunning {Effect of finite disk-thickness on swing amplification}

   \author{S. Ghosh\inst{1,2}
          \and C. J. Jog \inst{2}}

	\institute{ Inter-University Centre for Astronomy and Astrophysics, Pune 411007, India\\\email{soumavo@iucaa.in}
              \and
            Department of Physics, Indian Institute of Science, Bangalore 560012, India\\
          \email{cjjog@iisc.ac.in} }
              


 
 \abstract
 {A typical galactic disk is observed to have a finite thickness. Here, we present the study of the physical effect of introduction of finite thickness on the generation of small-scale spiral arms by swing amplification in a differentially rotating galactic disk. The galactic disk is modelled first as a one-fluid system, and then as a gravitationally-coupled two-fluid (stars and gas) system where each fluid is taken as  isothermal, and corotating with each other. We derived the equations governing the evolution of growth of the non-axisymmetric perturbations in a sheared frame of reference while incorporating the  effect of finite thickness of a galactic disk. We found that the finite thickness of a galactic disk has a generic trend of suppressing the growth of the non-axisymmetric perturbations via swing amplification. Moreover, even the observed range of disk-thickness values ($\sim$ 300-500 pc) can lead to a complete suppression of swing amplification for $Q$ $\sim$ 1.7, whereas for an infinitesimally-thin disk, the corresponding critical value is $Q \sim 2$. For a two-fluid (stars and gas) system, the net amplification is shown to be set by the mutual interplay of the effect of interstellar gas in promoting the spiral features and the effect of finite thickness in preventing the spiral arms. The coexistence of these two opposite effects is shown to be capable of giving rise to diverse and complex dynamical behaviour.

 }

   \maketitle
%

\section{Introduction} 

Several past studies, starting from \citet{GLB65,JT66,Too81} have shown conclusively that a galactic disk responds remarkably to the non-axisymmetric perturbations even though the disk is stable against the axisymmetric perturbation. This finding holds true whether the galactic disk is modelled as collisionless \citep{JT66,Too81} or as fluid \citep{GLB65}. It was shown that due to the mutual interplay among the shear of disk, epicyclic motion of the particles and the self-gravity of the disk, the initial non-axisymmetric perturbations get amplified, grow for a limited time, before finally being smeared out by the shear of the system. This phenomenon was identified as the mechanism for producing small-scale spiral features (material arms) in the disk. The `swing amplification' process as coined by \citet{Too81}, continued to be explored in later times by means of analytical methods \citep[e.g. see][]{Ath84,Jog92,Fu01,MK16a,MK16b} and $N$-body simulations \citep[e.g. see][]{SC84,CF85,Sel11,Fuj11,BSW13,Don13,Gra13}, along with some applications to external galaxies \citep[e.g. see][]{GJ14,Don15,GJ18}.

Although these studies, taken together, have revealed many useful insights about the generation of spiral features and their subsequent effect on the secular evolution of the galactic disk, however these studies assumed a number of approximations. For example, most of the analytical studies assumed an infinitesimally-thin disk for the simplicity of the calculation \citep[however see][]{GLB65}. This assumption is reasonably justified when the height of the disk is small compared to the wavelength of the perturbation \citep[e.g. see][]{Too64}.

In reality, a galactic disk is never infinitesimally-thin by its nature. For example, the thin-disk component of Milky Way has a scale-height of $\sim$ 300 pc at the solar neighbourhood \citep[e.g. see][]{Jur08}, and therefore, it would be worthwhile to investigate what the effects of the finite disk would be. This is the motivation for our work. Further, the height of the thin stellar disk is known to increase with radius, by a factor of 2--3 within the optical radius, as seen observationally \citep{deg97} and also shown on general theoretical grounds \citep{NJ02b}. Recent $N$-body simulations by \citet{Kaw17} showed that an initially flared, thin-disk model can produce a negative vertical metallicity gradient (due to radial mixing driven by bars and spiral arms), consistent with the current observational trend. Recent observations of our Galaxy \citep{Lop14} have shown the scale-height to increase substantially between radii from 8.5 kpc to 25 kpc. Thus, all these facts considered together, indicate that a more realistic study of small-scale spiral structure generated by the swing amplification process should include the effect of the finite thickness of a galactic disk.

 In the literature, it has been shown that the introduction of  finite thickness of a galactic disk leads to a reduction in the radial force in the mid-plane and can be treated as an effective reduction in the surface density in the mid-plane \citep[e.g. see][]{Too64,JS84,Jog14}. This in turn will tend to make the galactic disk more stable against the axisymmetric perturbations \citep{Too64,Jog14}. \citet{GLB65} derived the equations for evolution of non-axisymmetric perturbations in a sheared frame for a fluid disk with finite thickness, and \citet{JT66} showed that for a local patch of a stellar disk, the finite thickness reduces the amplitude of the density transforms. However, the effect of the finite thickness of a galactic disk on the growth of non-axisymmetric perturbations via
swing amplification and hence on the the small-scale spiral arms has not been explored systematically so far.

In this paper, we revisit the finite thickness problem for a galactic disk modelled as fluid and investigate the physical effect of finite thickness of a galactic disk on the resulting swing amplification process by employing a much wider ranges of input parameters (such as Toomre $Q$ parameter). The usage of wider ranges of input parameters will help us to understand the physical effect of the finite thickness on swing amplification in detail. 

We modelled the galactic disk first as a one-fluid system and then as a gravitationally-coupled two-fluid (stars and gas) system where the gas has a lower velocity dispersion as compared to stars. We derived the equations describing the evolution of local, non-axisymmetric perturbations in sheared coordinates for both the one-fluid and two-fluid systems having finite thickness. Using these equations, we investigated the effect of finite thickness on the efficiency of the swing amplification mechanism. Since a mode with highest Maximum Amplification Factor (MAF) (for definition see \S~2.2) is likely to stand out in a real system out of all possible modes, therefore we have focused on how the MAF changes as a function of increasing thickness of the disk, first for a one-fluid galactic disk and then for a two-fluid system.

We show that the inclusion of finite thickness of a galactic disk decreases the MAF of the resulting swing-amplified features in the disk plane. More than that, for some ranges of thickness values (which lie well within the observed ranges of thickness of the disks), the swing amplification is damped almost completely. This holds true for wide ranges of Toomre $Q$ parameters and thickness of the disk, and hence the effect is generic.

 We find that for a gravitationally-coupled two-fluid (star-gas) system,  the mutual interplay between the effect of interstellar gas in helping to host strong spiral features and the effect of finite thickness in preventing the strong spiral features is capable of showing a range of diverse and physically rich scenarios which otherwise can not be obtained from infinitesimally-thin modelling of a galactic disk.

The rest of the paper is organized as follows. \S~2 gives the derivation of the equations for swing amplification for a galactic disk having finite thickness. \S~3 describes the results and \S~4 discusses some applications for realistic galaxies while \S~5 and \S~6 contain the discussion and conclusions, respectively.

\section{Formulation of the problem}

Following the formulation of \citet{Jog92}, first we started with an infinitesimally thin galactic disk, and subsequently we incorporated the effect of the finite thickness of a galactic disk in the equations.

\subsection{Non-axisymmetric perturbation in infinitesimally thin fluid disk}

 The formulation of local, non-axisymmetric linear perturbation analysis of a galactic disk is largely followed from \citet{Jog92}. For the sake of completeness, here we only mention the relevant assumptions and equations, for details see \citet{Jog92}.

The baryonic component (stars or gas) in the galactic disk is modelled as an isothermal fluid, characterized by the surface density $\Sigma$ and the one-dimensional velocity dispersion or the sound speed $c$. Next we performed the linear perturbation analysis on the Euler's equations of motion, the continuity equation, and the Poisson equation in a sheared coordinate system defined as:
\begin{equation}
x'=x,\: y'=y-2Axt,\: z'=z, \: t'=t\,.
\end{equation}

We define $\tau$ as:\\
\begin{equation}
\tau \equiv 2At'-k_x/k_y \hspace{0.3 cm} \mbox{, for a wavenumber} \hspace{0.1 cm} k_y \ne 0
\end{equation}
In the sheared coordinates, $\tau$ is a measure of time, and it becomes zero when the modes becomes radial, that is, where $x$ is along the initial radial direction.

Now, assuming a trial solution of the form exp[$i(k_x x' + k_yy')$] for the independent perturbed quantities, for example, perturbed surface density $\delta \Sigma$, the local, linearized perturbed equations of motion, continuity equation and the Poisson equation become \citep[for details see][]{Jog92}: 

\begin{equation}
\frac{\partial v_x}{\partial \tau}-\frac{\Omega}{A}v_y=- i\frac{k_y}{2A}\tau \left[-\delta \Phi - \frac{c^2}{\Sigma_0}(\delta \Sigma)\right]\,,
\end{equation}

\begin{equation}
\frac{\partial v_y}{\partial \tau}+\frac{B}{A}v_x= i\frac{k_y}{2A} \left[-\delta \Phi - \frac{c^2}{\Sigma_0}(\delta \Sigma)\right]\,,
\end{equation}

\begin{equation}
\frac{\partial}{\partial \tau}(\delta \Sigma)-i\frac{k_y}{2A}\tau \Sigma_0 v_x+ i\frac{k_y}{2A}\Sigma_0 v_y=0\,,
\end{equation}
\noindent and,

\begin{equation}
\left[-k^2_y(1+\tau^2)-\frac{\partial'^2}{\partial z^2}\right](\delta \Phi)=4\pi G \delta \Sigma \delta(z')\,,
\label{eq-per-poi}
\end{equation}
\noindent respectively. Here, $v_x$ and $v_y$ are the perturbed velocity components in the $x$ and $y$ directions, respectively, and $A$, $B$ are the Oort constants. $\Sigma_0$ denotes the unperturbed surface density, and $\Omega$ is the circular velocity at radius $R$.

We note that, the wavenumber is constant in the sheared coordinate system whereas in the uniformly rotating frame it increases with $\tau$ ($k_{\rm non-sheared}=k_y[1+\tau^2]^{1/2}=k_y \sec\gamma'$) \citep[for details see][]{Jog92}. Now, for an infinitesimally thin disk, the solution of the perturbed Poisson equation (Eq.~\ref{eq-per-poi}) becomes \citep[for details see e.g.][]{GLB65,Jog92}

\begin{equation}
\delta \Phi =\left[-\frac{2\pi G}{k_y(1+\tau^2)^{1/2}}\right](\delta \Sigma)\,,
\label{eq-sol-poi}
\end{equation}
\noindent where $\delta \Phi$ is the perturbed potential.

\subsection{Introduction of finite thickness in the formulation}

For simplicity, we assume that the disk has a constant density that does not vary with $z$, and the disk has a total thickness of $2h$. Now the finite thickness of the disk leads to a reduction in the radial force in the mid-plane ($z = 0$) by a factor $(1-exp(-kh))/kh$ in the axisymmetric case where $k$ is the wavenumber of the perturbation \citep[for details see e.g.][]{Too64,JS84,Jog14}. This can be thought of as a reduction in the disk surface density \citep{Too64}. 

In an analogous way, for the non-axisymmetric case, the finite thickness 
of the disk introduces a reduction in the perturbed force terms in the Euler 
equations along $x'$ and $y'$ directions in the sheared frame. The form of that
reduction factor, say $\delta$, is given by

\begin{equation}
\delta = \left[\frac{1-exp(-k_y (1+\tau^2)^{1/2}h)}{k_y (1+\tau^2)^{1/2}h}\right]\,,
\label{eq-sol-poi-finite}
\end{equation}

Now we define $\theta$, the dimensionless measure of the density perturbation as:\\
\begin{equation}
\theta = \delta \Sigma/ \Sigma\,.
\end{equation}

Here, $\delta \Sigma$ denotes the variation in surface density in the sheared frame whereas in the non-sheared galactocentric frame of reference, it denotes the density for a mode of wavenumber $k_y(1+\tau^2)^{1/2}$ that is sheared by an angle $\gamma'$ (= tan$^{-1}\tau$) with respect to the radial position. It implies that the higher the value of $\tau$ is, the more sheared will be the mode. 

Next, we solved the local, linearized perturbed Euler equations, continuity 
equation and the Poisson equation (Eqns. (3)-(5), and (7)) following the 
procedure as given in \citet[see Eqns. 41-72 there]{GLB65}. We find
that the reduction factor in the force terms leads to a modification only in 
the self-gravity term (which contains $\delta$ times the surface density) in 
the net equation that gives variation in $\theta$ with $\tau$. Therefore, 
$\delta$ can be thought of as an effective reduction in the surface density 
in a similar manner as shown by \citet{Too64} for the axisymmetric case. 
The resulting  equation that gives the variation of $\theta$ with $\tau$ 
is thus obtained to be

\begin{equation}
\begin{split}
\left(\frac{d^2\theta}{d\tau^2}\right)-\left(\frac{d\theta}{d\tau}\right)\left(\frac{2\tau}{1+\tau^2}\right)+\theta\Bigg[\frac{\kappa^2}{4A^2}+\frac{2B/A}{1+\tau^2}
+\frac{k_y^2 c^2}{4A^2}(1+\tau^2)\\
-\Sigma\left(\frac{\pi G k_y}{2A^2}\right)(1+\tau^2)^{1/2}\frac{1-exp(-k_y (1+\tau^2)^{1/2}h)}{k_y (1+\tau^2)^{1/2}h}\Bigg]=0\,,
\label{swing-1comp}
\end{split}
\end{equation}
\noindent where $\kappa$ is the local epicyclic frequency. 

The four terms within the square bracket of Equation (\ref{swing-1comp}) represent the epicyclic motion, the unperturbed shear flow, the pressure of the fluid (stars or gas), and the self-gravity modified due to the finite thickness of the disk, respectively. For small  $\tau$ values, the epicyclic motion term and the unperturbed shear flow dominates over the pressure term and for a flat rotation curve they cancel each other completely. This results in setting up a kinematic resonance. In addition to that if the self-gravity term dominates over the pressure term then the duration of kinematic resonance increases and the mode undergoes a swing amplification while evolving from radial position ($\tau =0$) to trailing position ($\tau > 0$) \citep[for details see][]{GLB65,Too81}. However, for large $|\tau|$ values , the pressure term dominates over other terms, and the corresponding solution will be oscillatory in nature.
 
   We define maximum amplification factor (MAF) for a given mode as follows :
\begin{equation}
MAF \equiv (\theta)_{\rm max}/(\theta)_{\rm ini}\,,
\end{equation}
\noindent where $\theta_{\rm max}$ and $\theta_{\rm ini}$ are the maximum and initial amplitudes of the oscillation, respectively.

Now we introduce three dimensionless parameters, namely, Toomre $Q$ parameter \citep{Too64} = $\kappa c / \pi G \Sigma$, $\eta$ (= $2A/\Omega$) which denotes the logarithmic shearing rate and $X$ = ($\lambda_y/\lambda_{\rm crit}$), where $\lambda_{\rm crit}= 4\pi^2 G \Sigma/\kappa^2$.
Also, we express the quantity $k_yh$ as \\
$k_y h=k_y/k_{\rm crit}\times (k_{\rm crit}h) = X^{-1}\beta$, where $\beta = k_{\rm crit}h$.

Equation~(\ref{swing-1comp}) can then be written in terms of these quantities, which in turn gives the evolution of $\theta$ with $\tau$ as:
\begin{equation}
\begin{split}
\left(\frac{d^2\theta}{d \tau^2}
\right)-\left(\frac{d\theta}{d\tau}\right)\left(\frac{2\tau}{1+\tau^2}
\right)+\theta\Bigg[{\xi^2}+\frac{2(\eta-2)}{\eta(1+\tau^2)}
+\frac{(1+\tau^2)Q^2\xi^2}{4X^2}\\
-\frac{\xi^2}{X}(1+\tau^2)^{1/2}\frac{1-exp\{-X^{-1}\beta(1+\tau^2)^{1/2}\}}{X^{-1} \beta (1+\tau^2)^{1/2}}\Bigg]=0\,,
\end{split}
\label{swing-one-final}
\end{equation}
\noindent where ${\xi^2}={\kappa^2/4A^2}= 2(2-\eta)/\eta^2$.

It is straightforward to check that in the limit $h \rightarrow 0$, Equation~(\ref{swing-one-final}) reduces to the standard equation for an infinitesimally thin disk which is given as \citep[e.g. see][]{Jog92,GJ14}
\begin{equation}
\begin{split}
\left(\frac{d^2\theta}{d \tau^2}
\right)-\left(\frac{d\theta}{d\tau}\right)\left(\frac{2\tau}{1+\tau^2}
\right)+\theta\Bigg[{\xi^2}+\frac{2(\eta-2)}{\eta(1+\tau^2)}
+\frac{(1+\tau^2)Q^2\xi^2}{4X^2}\\
-\frac{\xi^2}{X}(1+\tau^2)^{1/2}\Bigg]=0\,.
\end{split}
\label{swing-final-infinite}
\end{equation}

For a given set of parameter values we solve Equation~(\ref{swing-one-final}) numerically by fourth-order Runge-Kutta method while treating Equation~(\ref{swing-one-final}) as two coupled, first--order linear differential equations in $\theta$ and $d\theta/d\tau$. 

\subsection{Swing amplification in two-fluid model of galactic disk with finite thickness}

Here in this section, we briefly present the derivation of the equations for the swing amplification in a galactic disk where the galactic disk is modelled as a gravitationally-coupled two-fluid (stars and gas) system, with each component corotating with each other. Each component is assumed to be isothermal, and they are characterized by surface density $\Sigma_i$, one-dimensional velocity dispersion or the sound speed $c_{i}$, where $i =s, g$ denotes the stars and gas, respectively. The stellar velocity dispersion is assumed to be higher than that of gas, in accordance with the observed trends seen in galaxies in the local Universe.

\subsubsection{One component has finite thickness, the other is infinitesimally-thin}

First, we modelled the galactic disk as consisting of two isothermal fluids where one component (stars) has a total finite thickness of $2h_1$ and the other component (gas) is infinitesimally-thin.  This particular modelling of a galactic disk, although somewhat contrived, allows us to investigate the effect of finite thickness of one fluid on the resulting swing amplification of the other fluid which is taken as infinitesimally-thin in nature.

Following the procedure as given in \S~2.1, the the local, linearized perturbed equations of motion, continuity equation, and Poisson equation in the sheared frame can be expressed as

\begin{equation}
\frac{\partial v_{x_i}}{\partial \tau}-\frac{\Omega}{A}v_{y_i}=- i\frac{k_y}{2A}\tau \left[-\delta \Phi_{\rm tot} - \frac{c_{i}^2}{\Sigma_0}(\delta \Sigma_i)\right]\,,
\end{equation}

\begin{equation}
\frac{\partial v_{y_i}}{\partial \tau}+\frac{B}{A}v_{x_i}= i\frac{k_y}{2A} \left[-\delta \Phi_{\rm tot} - \frac{c_{i}^2}{\Sigma_0}(\delta \Sigma_i)\right]\,,
\end{equation}

\begin{equation}
\frac{\partial}{\partial \tau}(\delta \Sigma_i)-i\frac{k_y}{2A}\tau \Sigma_{0i} v_{x_i}+ i\frac{k_y}{2A}\Sigma_{0i} v_{y_i}=0\,,
\end{equation}

\noindent and,

\begin{equation}
\delta \Phi_{\rm tot} =\left[-\frac{2\pi G}{k_y(1+\tau^2)^{1/2}}\right](\delta \Sigma_{\rm s} + \delta \Sigma_{\rm g})\,,
\label{eq-sol-poi2}
\end{equation}
\noindent respectively. $\delta \Phi_{\rm tot}$ (= $\delta \Phi_{\rm s} +  \delta \Phi_{\rm g}$) is the total perturbed gravitational potential. $\Sigma_{0i}$ and $\delta \Sigma_i$ are the unperturbed and perturbed surface densities for the $i^{th}$ component, respectively.

In analogy to the one-fluid finite thickness case, for the non-axisymmetric two-fluid case also where one fluid (stars) has finite thickness and the other (gas) being infinitesimally-thin, the perturbed forced terms in the Euler equations for the fluid
(having finite thickness) will have a reduction factor, $\delta_{\rm s}$ whose form is given as

\begin{equation}
\delta_{\rm s} = \left(\frac{1-exp(-k_y (1+\tau^2)^{1/2}h_1)}{k_y (1+\tau^2)^{1/2} h_1}\right)\,.
\label{eq-sol-two-poi-semi_finite}
\end{equation}

Now in a similar way as done for the one-fluid case, we find that the 
reduction factor in the force terms results in a modification only in the
self-gravity term (which contains $\delta$ times the surface density) 
in the net equation that gives the variation in $\theta$ with $\tau$. 
Therefore, $\delta$ can be thought of as an effective reduction in the 
corresponding surface density. The net coupled equations 
that give the evolution of $\theta_i$ with $\tau$ are obtained to be

\begin{equation}
\begin{split}
\left(\frac{d^2\theta_i}{d\tau^2}\right)-\left(\frac{d\theta_i}{d\tau}\right)\left(\frac{2\tau}{1+\tau^2}\right)+\theta_i\Bigg[\frac{\kappa^2}{4A^2}+\frac{2B/A}{1+\tau^2}
+\frac{k_y^2c_{i}^2}{4A^2}(1+\tau^2)\Bigg]\\
=\left(\frac{\pi G k_y}{2A^2}\right)(1+\tau^2)^{1/2} \left[\delta \Sigma_{\rm s}\left(\frac{1-exp(-k_y (1+\tau^2)^{1/2}h_1)}{k_y (1+\tau^2)^{1/2} h_1}\right)+\delta \Sigma_{\rm g}\right]\,,
\end{split}
\end{equation}
\noindent where $i=$ s, g for stars and gas, respectively.

As before, we introduce dimensional quantities such as Toomre $Q$ parameter ($Q_i= \kappa c_{i}/\pi G \Sigma_i$), mass fraction ($\epsilon$) = $\Sigma_{\rm g}/(\Sigma_{\rm s}+\Sigma_{\rm g})$, $\eta$ (= $2A/\Omega$), $X$ = ($\lambda_y/\lambda_{\rm crit}$), where $\lambda_{\rm crit}= 4\pi^2 G (\Sigma_{\rm s}+\Sigma_{\rm g})/\kappa^2$, and $\beta_1$ $= k_{\rm crit}h_1$. In terms of these dimensionless quantities, the final equations expressing the evolution of $\theta_i$ with $\tau$ reduce to

\begin{equation}
\begin{split}
\left(\frac{d^2\theta_{\rm s}}{d \tau^2}
\right)-\left(\frac{d\theta_{\rm s}}{d\tau}\right)\left(\frac{2\tau}{1+\tau^2}
\right)\\
+\theta_{\rm s}\Bigg[{\xi^2}+\frac{2(\eta-2)}{\eta(1+\tau^2)}
+\frac{(1+\tau^2)Q_{\rm s}^2(1-\epsilon)^2\xi^2}{4X^2}\Bigg]\\
=\frac{\xi^2}{X}(1+\tau^2)^{1/2}\Bigg[\theta_{\rm s}(1-\epsilon)\frac{1-exp(-X^{-1}\beta_1(1+\tau^2)^{1/2})}{X^{-1} \beta_1(1+\tau^2)^{1/2}}+\theta_{\rm g}\epsilon\Bigg]\,,
\end{split}
\label{swing2semi_one}
\end{equation}

\noindent and,

\begin{equation}
\begin{split}
\left(\frac{d^2\theta_{\rm g}}{d \tau^2}
\right)-\left(\frac{d\theta_{\rm g}}{d\tau}\right)\left(\frac{2\tau}{1+\tau^2}
\right)\\
+\theta_{\rm g}\Bigg[{\xi^2}+\frac{2(\eta-2)}{\eta(1+\tau^2)}
+\frac{(1+\tau^2)Q_{\rm g}^2 \epsilon^2\xi^2}{4X^2}\Bigg]\\
=\frac{\xi^2}{X}(1+\tau^2)^{1/2}\Bigg[\theta_{\rm s}(1-\epsilon)\frac{1-exp(-X^{-1}\beta_1(1+\tau^2)^{1/2})}{X^{-1} \beta_1(1+\tau^2)^{1/2}}+\theta_{\rm g}\epsilon\Bigg]\,.
\end{split}
\label{swing2semi_two}
\end{equation}

The maximum amplification factor (MAF) for each component is defined separately as in Equation~(11).

From Equations (\ref{swing2semi_one}) and (\ref{swing2semi_two}) it is evident that the solution for any individual fluid is governed by the self-gravity term set by both the fluid components, and hence introduction of finite thickness of a component is likely to have impact on the other fluid although the latter is modelled as infinitesimally-thin. In other words, for a gravitationally coupled two-fluid system, both the resulting $(\theta_i)$ and $(\theta_i)_{\rm max}$ for the infinitesimally-thin fluid (gas) will be affected by the finite thickness of the other fluid (stars), for results see \S~3.3.1. 

For a given set of parameter values we solved Equations (\ref{swing2semi_one}) and (\ref{swing2semi_two}) numerically by forth-order Runge-Kutta method while treating  Equations (\ref{swing2semi_one}) and (\ref{swing2semi_two}) as four coupled, first-order linear differential equations in $\theta_i$ and $d\theta_i/d\tau$. We point out that for this linear analysis, the ratio of perturbed to unperturbed surface density ($\theta_i$) may be multiplied by an arbitrary scale factor (say $\alpha$) such that the net fractional amplitude $\alpha \theta_i$ remains $\ll 1$, for all $\tau$ values considered \citep[for details see][]{Jog92}.

The corresponding condition for this model of galactic disk to be stable against the local, axisymmetric perturbation is \citep[for details see][]{JS84}

\begin{equation}
\begin{split}
\frac{(1-\epsilon)}{X'\{1+[Q_{\rm s}^2(1-\epsilon)^2/4X'^2]\}} \Bigg\{\frac{1-exp(-X'^{-1}\beta_1)}{X'^{-1}\beta_1}\Bigg\}\\
+ \frac{\epsilon}{X'\{1+[Q_{\rm g}^2 \epsilon^2/4X'^2]\}}<1\,,
\label{cond-axisymmetric_semi}
\end{split}
\end{equation}
\noindent where $X'= \lambda_a/\lambda_{\rm crit}$, and $\lambda_a$ denotes the wavelength of the axisymmetric perturbation. This ensures that the system is stable against axisymmetric perturbations, hence essentially we are solely dealing with the growth of the non-axisymmetric perturbations.

For a given set of parameters, the inequality (Eq.~\ref{cond-axisymmetric_semi}) has to be satisfied for all wavelengths ($X'$) between $Q_{\rm g}^2\epsilon^2/2$ to $Q_{\rm s}^2(1-\epsilon)^2/2$ \citep[for details see][]{JS84}. Also, while choosing a  set of parameters, we ensured that for that set of parameters, the ratio of gas velocity dispersion to stellar velocity dispersion is always less than unity \citep[for details see Eq. (31) in][]{Jog92}.

\subsubsection{Both components having finite thickness}

Here we employed a more realistic modelling of a galactic disk by treating it as a two-component  system consisting of two isothermal fluids (stars and gas) and each of the fluids has a  total thickness of $2h_i$, where $i=$ s, g for stars and gas, respectively.

Now it is straightforward to show that for such a two-fluid model with each 
component having a total thickness 2$h_i$, the net reduction in the perturbed
force terms in the Euler's equation in the sheared frame will have a reduction 
factor $\delta_i$ whose form is given by

\begin{equation}
\delta_i = \left(\frac{1-exp(-k_y (1+\tau^2)^{1/2}h_i)}{k_y (1+\tau^2)^{1/2} h_i}\right)\,.
\label{eq-sol-two-poi-finite}
\end{equation}

Next, in a similar way as done in \S~2.3.1, we find in the two-fluid case where each of the fluids has a total finite thickness of $2h_i$, the net equation that gives the variation in $\theta_i$ with $\tau$, only the surface density terms of the fluids having finite thickness are modified to contain the corresponding reduction factors $\delta_i$. Hence these can be thought of as the effective reduction factors for the corresponding surface densities. These coupled equations are obtained to be:
 
\begin{equation}
\begin{split}
\left(\frac{d^2\theta_i}{d\tau^2}\right)-\left(\frac{d\theta_i}{d\tau}\right)\left(\frac{2\tau}{1+\tau^2}\right)+\theta_i\Bigg[\frac{\kappa^2}{4A^2}+\frac{2B/A}{1+\tau^2}
+\frac{k_y^2c_{i}^2}{4A^2}(1+\tau^2)\Bigg]\\
=\left(\frac{\pi G k_y}{2A^2}\right)(1+\tau^2)^{1/2}\sum_{i} \delta \Sigma_i\frac{1-exp(-k_y(1+\tau^2)^{1/2} h_i)}{k_y(1+\tau^2)^{1/2} h_i}\,.
\label{swing-2comp}
\end{split}
\end{equation}

 In terms of the dimensionless quantities introduced above, the final equations expressing the evolution of $\theta_i$ with $\tau$ reduce to
 
\begin{equation}
\begin{split}
\left(\frac{d^2\theta_{\rm s}}{d \tau^2}
\right)-\left(\frac{d\theta_{\rm s}}{d\tau}\right)\left(\frac{2\tau}{1+\tau^2}
\right)\\
+\theta_{\rm s}\Bigg[{\xi^2}+\frac{2(\eta-2)}{\eta(1+\tau^2)}
+\frac{(1+\tau^2)Q_{\rm s}^2(1-\epsilon)^2\xi^2}{4X^2}\Bigg]\\
=\frac{\xi^2}{X}(1+\tau^2)^{1/2}\Bigg[\theta_{\rm s}(1-\epsilon)\frac{1-exp(-X^{-1}\beta_{\rm s}(1+\tau^2)^{1/2})}{X^{-1} \beta_{\rm s}(1+\tau^2)^{1/2}}+\\
\theta_{\rm g}\epsilon\frac{1-exp(-X^{-1}\beta_{\rm g}(1+\tau^2)^{1/2})}{X^{-1} \beta_{\rm g} (1+\tau^2)^{1/2}}\Bigg]\,,
\end{split}
\label{swing-21-final}
\end{equation}

\noindent and,

\begin{equation}
\begin{split}
\left(\frac{d^2\theta_{\rm g}}{d \tau^2}
\right)-\left(\frac{d\theta_{\rm g}}{d\tau}\right)\left(\frac{2\tau}{1+\tau^2}
\right)\\
+\theta_{\rm g}\Bigg[{\xi^2}+\frac{2(\eta-2)}{\eta(1+\tau^2)}
+\frac{(1+\tau^2)Q_{\rm g}^2 \epsilon^2\xi^2}{4X^2}\Bigg]\\
=\frac{\xi^2}{X}(1+\tau^2)^{1/2}\Bigg[\theta_{\rm s}(1-\epsilon)\frac{1-exp(-X^{-1}\beta_{\rm s}(1+\tau^2)^{1/2})}{X^{-1} \beta_{\rm s}(1+\tau^2)^{1/2}}\\
+\theta_{\rm g}\epsilon\frac{1-exp(-X^{-1}\beta_{\rm g}(1+\tau^2)^{1/2})}{X^{-1} \beta_{\rm g}(1+\tau^2)^{1/2}}\Bigg]\,.
\end{split}
\label{swing-22-final}
\end{equation}

We checked that, in the limits of $\beta_{\rm s} \rightarrow 0$ and $\beta_{\rm g} \rightarrow 0$, Equations (\ref{swing-21-final}) and (\ref{swing-22-final}) reduce to the standard equation for an infinitesimally thin two-fluid disk, as expected \citep[see Eqs. (32 - 33) in][]{Jog92}.

 The condition for the system to be stable against the axisymmetric perturbation is \citep[for details see][]{JS84}

\begin{equation}
\begin{split}
\frac{(1-\epsilon)}{X'\{1+[Q_{\rm s}^2(1-\epsilon)^2/4X'^2]\}} \Bigg\{\frac{1-exp(-X'^{-1}\beta_{\rm s})}{X'^{-1}\beta_{\rm s}}\Bigg\}\\
+ \frac{\epsilon}{X'\{1+[Q_{\rm g}^2 \epsilon^2/4X'^2]\}}
\Bigg\{\frac{1-exp(-X'^{-1}\beta_{\rm g})}{X'^{-1}\beta_2}\Bigg\} <1\,,
\label{cond-axisymmetric}
\end{split}
\end{equation}
\noindent where $X'= \lambda_a/\lambda_{\rm crit}$, and $\lambda_a$ denotes the wavelength of the axisymmetric perturbation.

\section{Results}

First, we investigated the effect of finite thickness in the swing amplification process for a one-fluid case, and then we examined the effect of finite thickness for a gravitationally-coupled two-fluid galactic disk.

\subsection {Choice of $k_{\rm crit}$ value}

It is clear from the definition of $\beta$ that its value is dependent on the value of $k_{\rm crit}$ (see \S~2.2 for details). Therefore, for the same thickness of a disk, the values of $\beta$ will be different depending on the values of $k_{\rm crit}$. Here, for the sake of uniformity, we chose $k_{\rm crit}$ = 1 kpc$^{-1}$. We note that in the solar neighbourhood, a $v_{\rm c}$ (rotation velocity) of $\sim$ 220 km s$^{-1}$ and $\Sigma$ $\sim$ 45 M$_{\odot}$ pc$^{-2}$ \citep[e.g. see][]{Mera98,NJ02a} will produce $k_{\rm crit}$ $\sim$ 1 kpc$^{-1}$, thus justifying our assumption of $k_{\rm crit}$ value as being reasonable. See Appendix : A for the details of choice of $\beta$ values.

\subsection {Effect of finite thickness for one-fluid galactic disk}

\begin{figure}
\centering
\includegraphics[height=2.3in, width=3.3in]{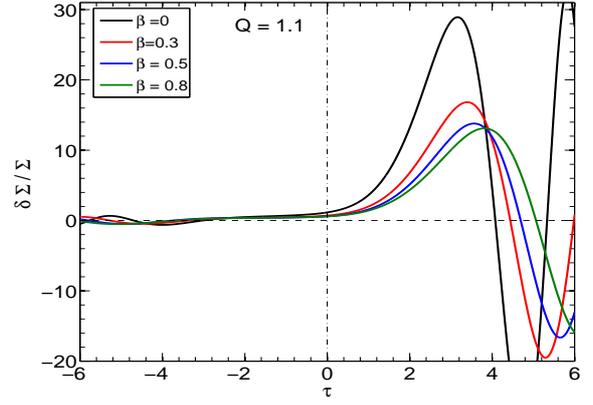}
\caption{Variation in $\theta =\delta\Sigma/\Sigma$, the ratio of the perturbed surface density to the unperturbed surface density, with $\tau$, dimensionless time in the sheared frame, plotted for three different $\beta$ values, and for $Q =1.1$, $X=1$, $\eta =1$, for a one-fluid case. As seen clearly, with the increase of the disk thickness, the resulting MAF of the solution decreases steadily. The net amplitude $\alpha \theta$ $\ll 1$at all $\tau$, where $\alpha$ is a scale factor. }
\label{fig:sample}
\end{figure}

We first examined the effect of finite thickness in a one-fluid galactic disk. We take  $Q = 1.1$, so that the disk is stable against the local, axisymmetric perturbation \citep{Too64} and still the self-gravity is important \citep[for details see][]{BT87}. Also, we have taken $X =1$, and $\eta =1$ (corresponding to a flat rotation curve). The corresponding solutions for different $\beta$ values are shown in Fig.~\ref{fig:sample}.

We note that we have taken $\eta =1$ throughout which corresponds to a `flat' rotation curve. However, the finite thickness only modifies the self-gravity term while leaving other terms unchanged (for details see \S~2.2), and hence the variation in $\eta$ will not have much effect on the findings presented here.

From Fig.~\ref{fig:sample}, two trends are evident which we mention here.\\
\begin{itemize}
\item{ The resulting MAF of the swing amplification decreases monotonically with the increase of thickness of the galactic disk. To express it quantitatively, the MAF for $\beta =0.8$ decreases by $\sim$ 53 \% as compared to what is seen for the infinitesimally-thin disk ($\beta =0$). This trend is in fair agreement with the results obtained for a stellar sheet with finite thickness as shown by \citet{JT66}.}

\item{The pitch angle, defined as $\gamma = tan^{-1}\tau_{\rm max}$ where $\tau_{\rm max}$ denotes the epoch of maximum amplification, changes only moderately. Quantitatively, it changes only $\sim$ 5 \% for a thickness of 800 pc ($\beta \sim 0.8$) as compared to an infinitesimally-thin disk.}

We note that the definition of pitch angle employed in this paper is different from the usual definition used in the literature \citep[e.g. see Fig. 6.8 in][]{BT87}. To compare our values of pitch angle with the standard usage, one has to take $i_{std} = 90^{0}-\gamma$, where $i_{std}$ is the pitch angle in the usual definition.
\end{itemize}

The physical explanation for these trends is as follows:\\
The introduction of finite thickness induces an effective reduction in the self-gravity term (for details see \S~2.2), and hence it can no longer dominate over the pressure term as strongly as it would have for the infinitesimally-thin case. During the rising phase or when the swing amplification sets in ($\tau \sim 0$), the self-gravity term has a weak dependence on $\tau$. Hence, the reduction factor due to disk thickness is effective for a larger $\tau$ range (or over several e-folding time) where the peak of the growth occurs, and therefore has a strong, non-linear damping effect on the resulting amplification \citep[see also][]{JT66}. The quantitative variation of the reduction factor with $\beta$ and $\tau$ is discussed later (see \S~3.3 \& 3.4).
As a result, the MAF decreases monotonically with the steady increase of the thickness of the disk. On the other hand, the pitch angle remains mostly unchanged since $\gamma = tan^{-1}\tau_{\rm max}$, by definition so it depends weakly on the values of $\tau_{\rm max}$. This is reflected in the marginal change in the pitch angle as a function of $\beta$.

Next we studied the systematic variation in the MAF of the resulting swing amplification as a function of $Q$ and finite thickness ($\beta$). The self-gravity term becomes progressively less important with the increase of Toomre $Q$ value \citep[e.g. see][]{BT87}, and the specific MAF of the swing-amplified features also decrease with the increase of $Q$ \citep[e.g. see][]{Too81}. Therefore it would be worth checking the effect of finite thickness for a wide range of Toomre $Q$ values.

\begin{figure}
\centering
\includegraphics[height=2.3in, width=3.3in]{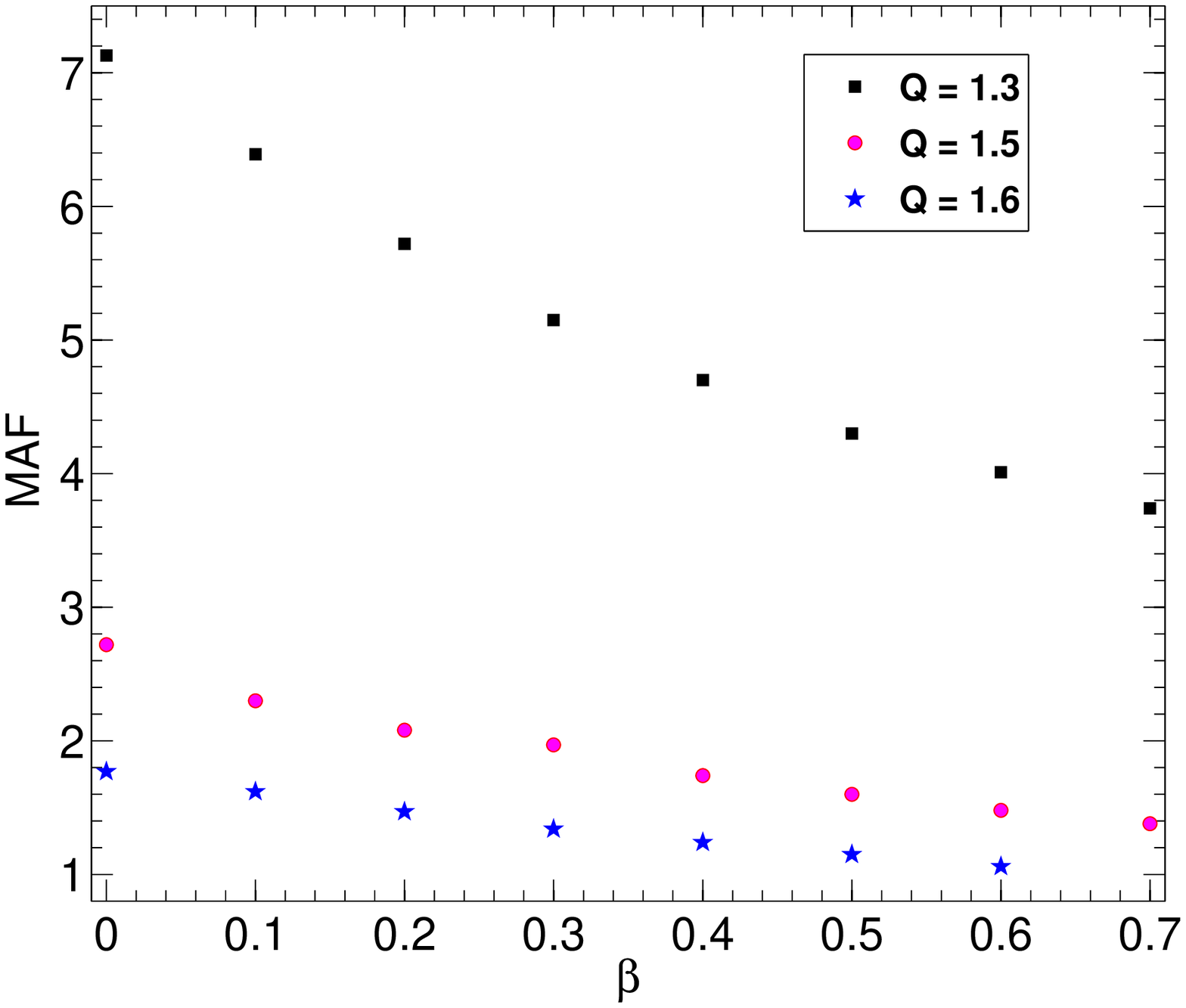}
\medskip
\includegraphics[height=2.3in, width=3.3in]{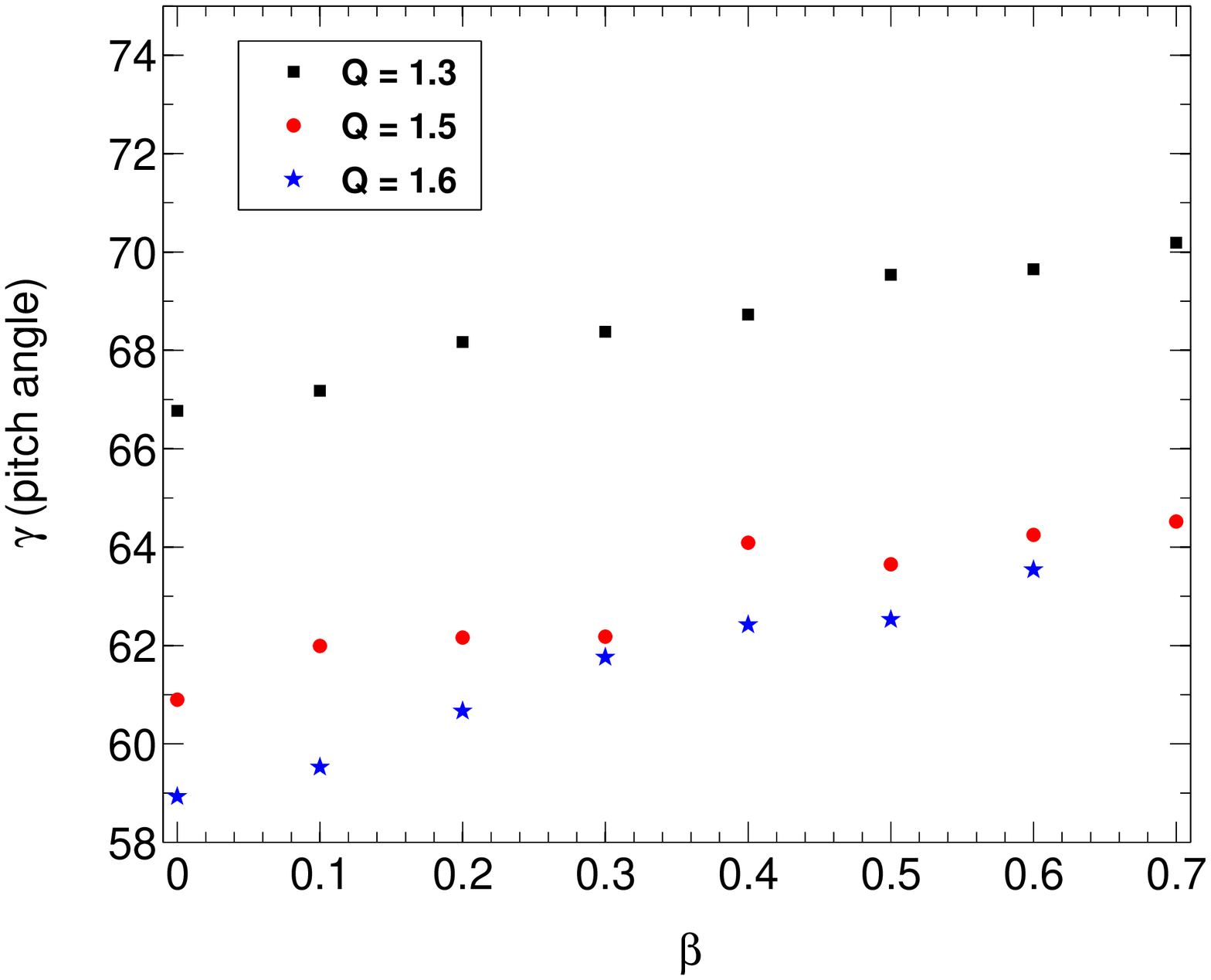}
\caption{Systematic variation of maximum amplification factor (MAF; for definition see \S~2.2) and the pitch angle of the resulting swing amplification, plotted as a function of disk scale-height ($\beta$) for different Toomre $Q$ values for a one-fluid case are shown in the top panel and bottom panel, respectively. For a given Toomre $Q$ value, the resulting MAF decreases steadily with the increase of disk thickness, whereas the pitch angle ($\gamma$) increases monotonically with the increase of disk thickness, thus implying the spiral features will be more tightly wound. The definition of pitch angle used here is different from the standard definition used in literature, for details see text. $90^o - \gamma$ will yield the pitch angle according to the standard definition. }
\label{fig:fig2}
\end{figure}

Fig.~\ref{fig:fig2} shows the systematic variation in the MAF as a function of finite thickness ($\beta$) for different $Q$ values. The MAF decreases monotonically with the steady increase of disk thickness, and this remains true for the whole range of Toomre $Q$ values we considered here. Quantitatively, for $Q =1.3$, the MAF is reduced by $\sim$ 48 \% for $\beta =0.7$ as compared to a infinitesimally-thin disk, whereas for $Q=1.5$, the change in MAF for $\beta=0.7$ is $\sim$ 50 \% as compared to the infinitesimally-thin disk. It means that the resulting swing amplified spiral features will be weaker than the case of infinitesimally-thin disk. Thus, introduction of finite thickness of the disk decreasing the amplitude of the resulting swing amplification turns out to be a generic trend, that is, it holds true for all Toomre $Q$ values. This is one of the main findings of the paper.

Interestingly, we find that finite thickness strongly affects the limiting value of Toomre $Q$ parameter denoting the total suppression of the growth of the non-axisymmetric perturbations. As shown in Fig.~\ref{fig:fig2}, for $Q=1.6$, any $\beta$ value more than $0.6$ (corresponding to a disk thickness of 600 pc) will prevent the swing amplification completely, thus the system will not be able to support any small-scale swing-amplified spiral features. We further checked that for even higher values of $Q$ (i.e. $Q \ge 1.7$) the system stops displaying swing amplification even for $\beta \ge 0.3$ (corresponding to a disk thickness of 300 pc, as typically seen in the galactic disk of our solar neighbourhood). In the past, $Q \ge 2$ was shown to be a sufficient condition for the stability against the non-axisymmetric perturbations \citep[e.g. see][]{Too81,CF85,Lar88}, while a smaller limit of $Q \ge 1.7$ was given by \citet{Pol89}. Here, we show that even $Q= 1.7$ and $\beta =0.3$ will be {\it sufficient} (as opposed to $Q=2$ for infinitesimally-thin case) to prevent the growth of the non-axisymmetric perturbations completely. Hence, the limiting Toomre $Q$ value denoting the complete suppression of growth of non-axisymmetric perturbation depend critically on the finite thickness of the disk. This is another main result of this paper.

\subsection {Effect of finite thickness for two-fluid galactic disk}

Here we investigated the effect of finite thickness on the resulting swing amplification in a galactic disk modelled as a gravitationally-coupled two-fluid (stars and gas) system. As we show in this section and later, the interstellar gas (having lower velocity dispersion than of stars) and the finite thickness have an opposite effect on the resulting swing amplification. Therefore, due to the gravitational coupling, the two-fluid model can show a diverse, and complex behaviour.

\subsubsection {Stellar disk has finite thickness, gas disk is infinitesimally-thin}

Here, in the two-fluid (star-gas) model for the galactic disk, the stellar disk has a total thickness of $2h_{\rm s}$, with the gas disk being treated as infinitesimally-thin. This allows us to isolate the effect of the finite thickness of the stellar disk on the swing amplification in the gas disk (treated as infinitesimally-thin).

\begin{figure}
\centering
\includegraphics[height=2.5in, width=3.5in]{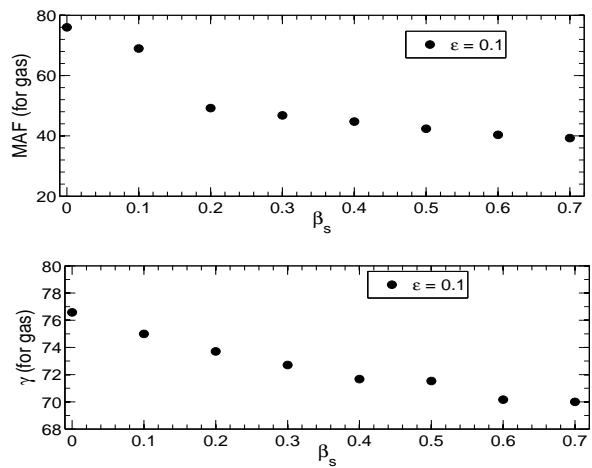}
\caption{Systematic variation of MAF (top panel; for definition see \S~2.3.1) and the pitch angle $\gamma$ (bottom panel) of the resulting swing amplification in the gas disk for a two-fluid (stars and gas) system where stellar disk has a finite thickness and the gas disk being infinitesimally-thin, plotted as a function of stellar disk finite thickness ($\beta_{\rm s}$). Here, $Q_{\rm s} = 1.5$, $Q_{\rm g} = 1.2$, $\epsilon = 0.1$, $\eta = 1$, $X =1$.  Although modelled as infinitesimally-thin, the MAF of the swing-amplified features in the gas disk decreases steadily with the increase of finite thickness (equivalently an increase in $\beta_{\rm s}$) of the other component due to the gravitational coupling between the two fluid components, for details see text.
 }
\label{fig:swing2semi_maf}
\end{figure}

First, we considered a case where $Q_{\rm s} = 1.5$, $Q_{\rm g} = 1.2$, $\epsilon =0.1$, $\eta= 1$ and $X=1$. This set of input parameters satisfies the inequality given by Equation (\ref{cond-axisymmetric_semi}), thus making the joint-system stable against the local, axisymmetric perturbation, and also at the same time allows finite swing amplification in both the components. We then systematically varied the finite thickness of the stellar disk from $\beta_{\rm s}=0.1$ to $\beta=0.7$, and for each case we have solved the Equations (\ref{swing2semi_one}) and (\ref{swing2semi_two}). The resulting MAFs and the pitch angles ($\gamma$) in the gas disk are shown in the Fig.~\ref{fig:swing2semi_maf}. Also, for comparison we calculated the resulting MAF for a infinitesimally-thin two-fluid case while keeping the input parameters unchanged (i.e. $Q_{\rm s} = 1.5$, $Q_{\rm g} = 1.2$, $\epsilon =0.1$, $\eta= 1$ and $X=1$).

From Fig.~\ref{fig:swing2semi_maf} it is clear that the MAF in the gas disk continues to decrease monotonically with the increase in the finite thickness of the stellar disk ($\beta_{\rm s}$). To state quantitatively, the MAF in the gas disk decreases by $\sim$ 48 \% for a case of $\beta_{\rm s} =0.7$ when compared against the MAF of the same gas disk for the infinitesimally-thin two-fluid case. Thus,  in spite of the fact that the gas disk is modelled as infinitesimally-thin and gas has a lower velocity dispersion (allowing larger growth of the non-axisymmetric perturbations), the finite thickness of the stellar disk affects the resulting MAF in the gas disk in a gravitationally-coupled system. This is one of the main findings of this paper. However, the pitch angle in the gas disk does not change appreciably and the change is only about $\sim$ 7-10 \%.

\begin{figure*}
    \centering
    \begin{minipage}{.32\textwidth}
        \centering
        \includegraphics[height=2.1in,width=2.3in]{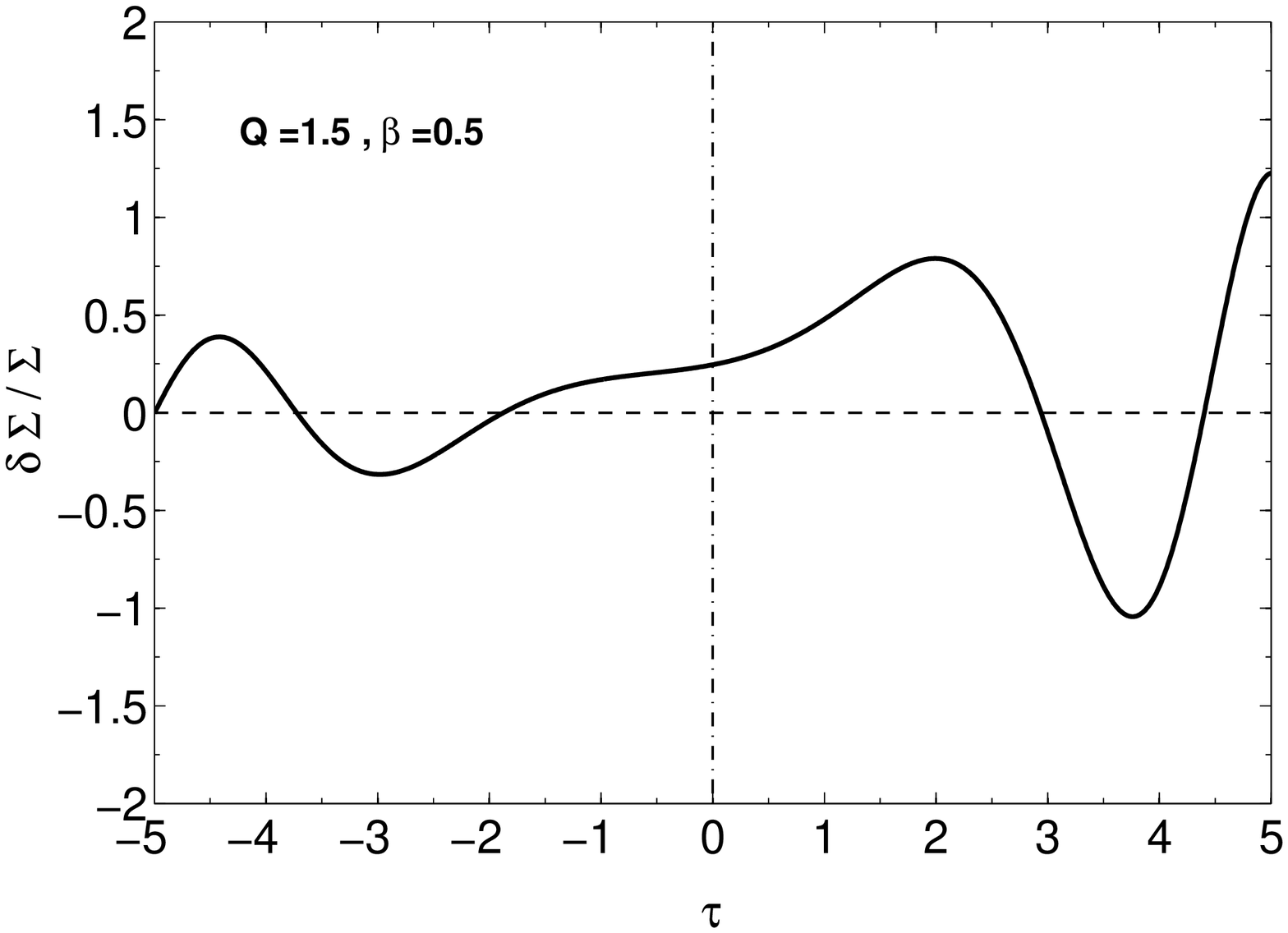}
       \vspace{0.2 cm}
	{\bf{(a)}}\\
    \end{minipage}
    \begin{minipage}{.32\textwidth}
        \centering
        \includegraphics[height=2.1in,width=2.3in]{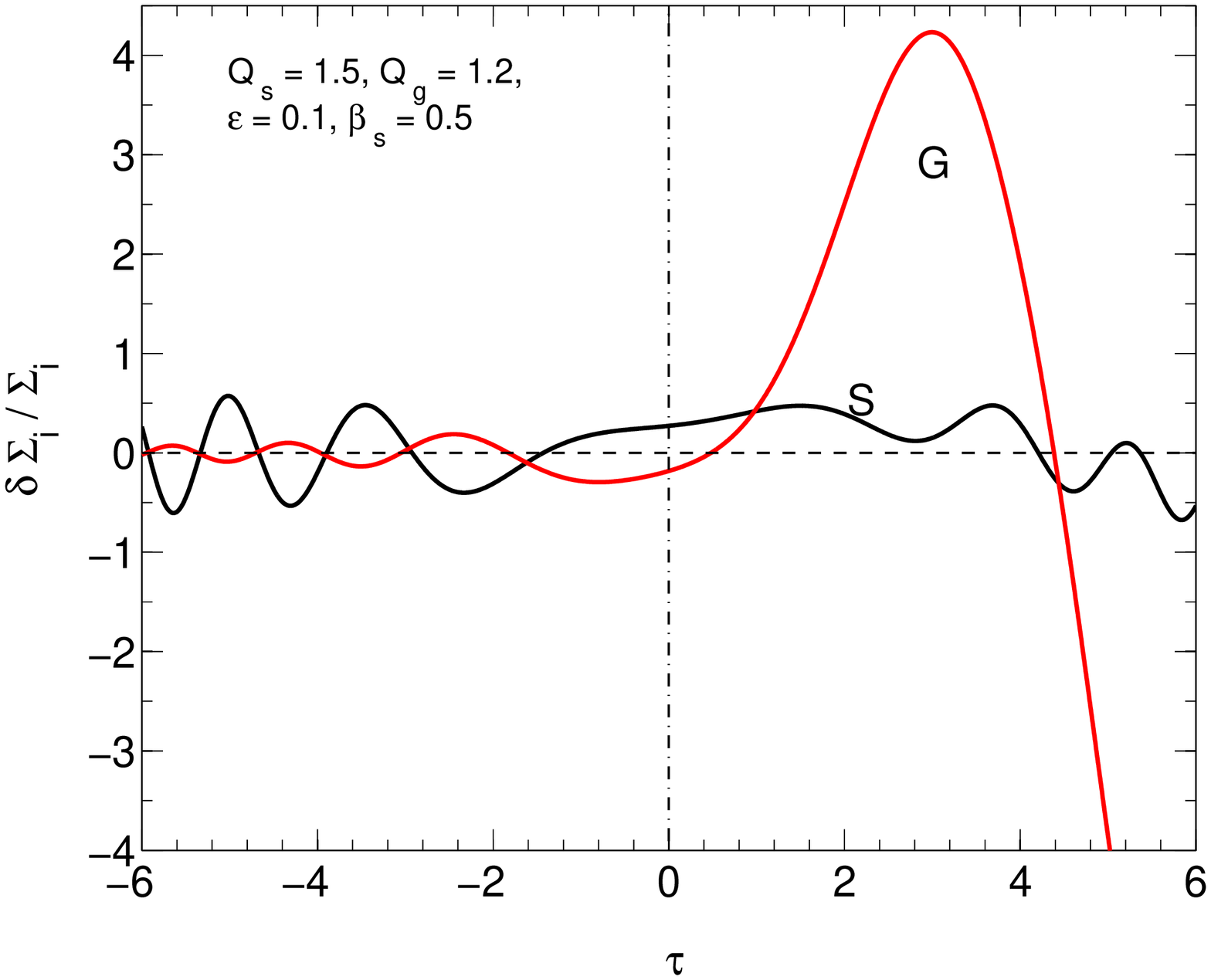}
        \vspace{0.2 cm}
	{\bf{(b)}}\\
    \end{minipage}
\begin{minipage}{.32\textwidth}
        \centering
        \includegraphics[height=2.1in,width=2.3in]{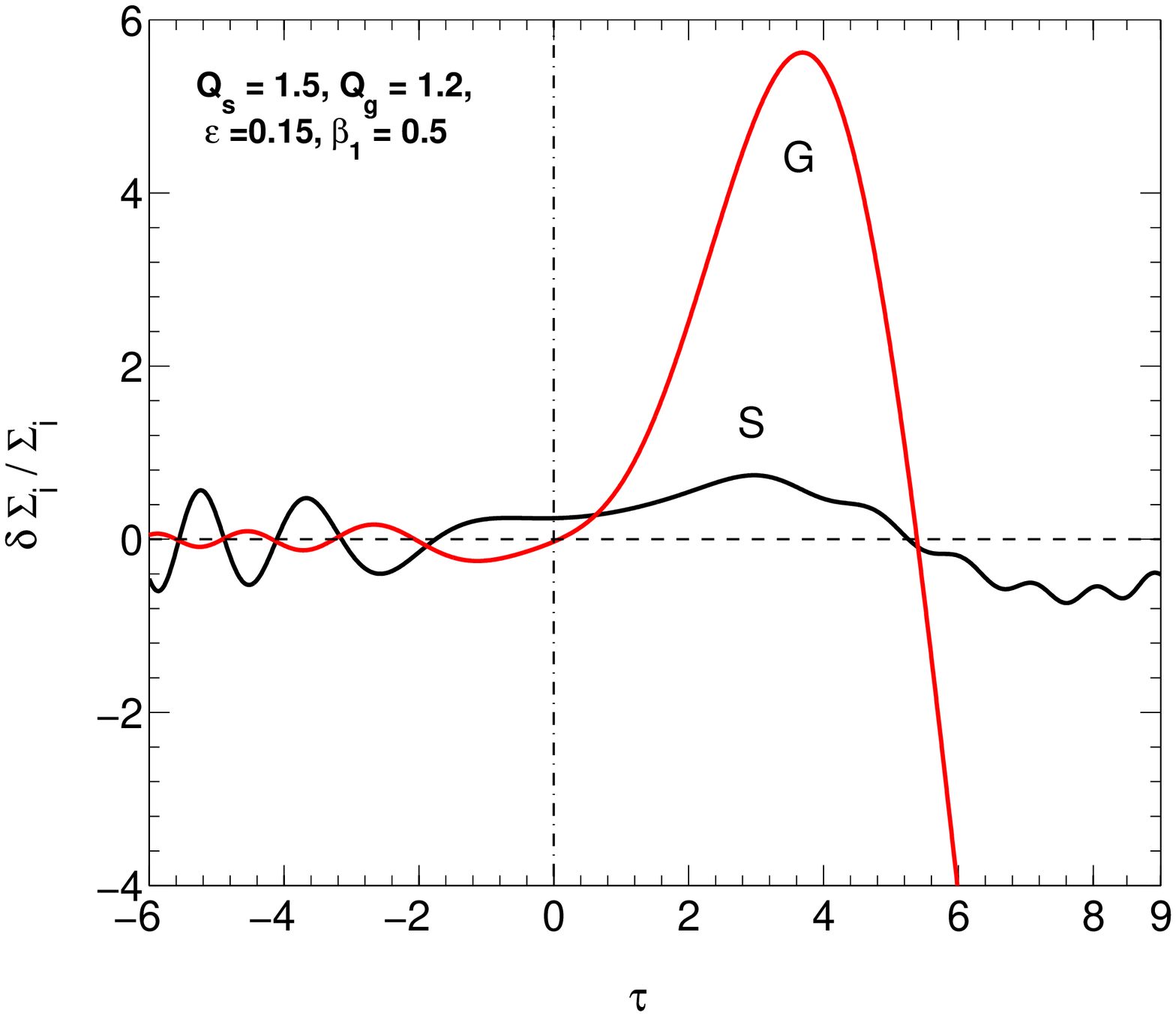}
        \vspace{0.2 cm}
	{\bf{(c)}}\\
    \end{minipage}
    \caption{Variation in $\theta =\delta\Sigma/\Sigma$, the ratio of the perturbation surface density to the unperturbed surface density, with $\tau$, dimensionless time in the sheared frame, plotted for three different models of a galactic disk.  (a) shows the case for a one-fluid system whereas (b) and (c) show results for gravitationally-coupled, two-fluid (stars and gas) cases where stellar disk has a finite thickness with the gas disk being infinitesimally-thin. The assumed input parameters ($Q_{\rm s}$, $Q_{\rm g}$, $\epsilon$, $\beta_{\rm s}$) are indicated in the legend. As seen clearly, the introduction of finite thickness reduces the MAF while gas increases the MAF, and hence the net amplification will be set by the dominant factor (for details see text). The net amplitude $\alpha \theta$ $\ll 1$at all $\tau$, where $\alpha$ is a scale factor.}
\label{fig:swing2semi_compare}
\end{figure*}

Also we checked that the MAF in the stellar disk decreases with the increase of finite thickness in the stellar disk, in agreement with the trend found for the one-fluid case (see \S~3.2). However, for this two-fluid model of galactic disk and the assumed parameter values, the stellar disk stops showing any finite amplification from $\beta_{\rm s} \ge 0.5$ (see Fig.~{\ref{fig:swing2semi_compare}} b), thus implying the absence of swing-amplified features in the stellar disk. This is surprising in the sense when the galactic disk is modelled as one-fluid system, the stellar disk would still display MAF grater than unity even for $\beta_{\rm s} \ge 0.5$ (see Figs.~{\ref{fig:swing2semi_compare}} a \& \ref{fig:fig2}).

To explain this, we note that in a galactic disk modelled as a gravitationally-coupled, two-fluid system, the low velocity dispersion component, namely, gas tends to increase the MAF of swing-amplified features in the stellar disk and the maximum amplification occurs at a later epoch (i.e. for larger $\tau$ values) even when the contribution of gas is moderate \citep[for details see][]{Jog92}. On the other hand, finite thickness of the disk tends to diminish the MAF of the swing-amplified features, as shown in the earlier sections. Therefore, the net MAF of the swing-amplified features in the stellar disk will be decided by the mutual interplay between these two opposite effects. 

\begin{figure*}
    \centering
    \begin{minipage}{.32\textwidth}
        \centering
         \includegraphics[height=2.1in,width=2.3in]{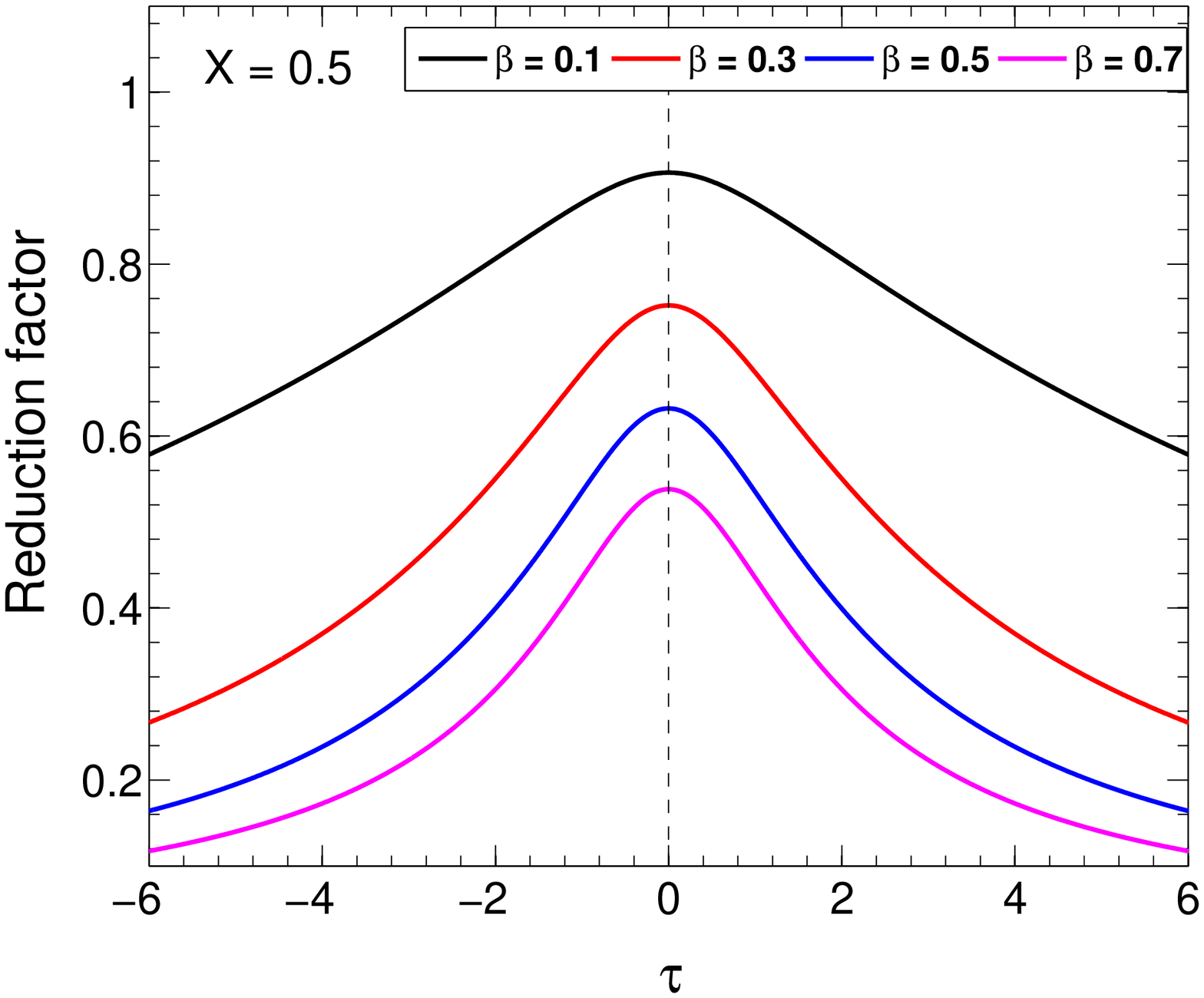}
       \vspace{0.2 cm}
	{\bf{(a)}}\\
    \end{minipage}
    \begin{minipage}{.32\textwidth}
        \centering
        \includegraphics[height=2.1in,width=2.3in]{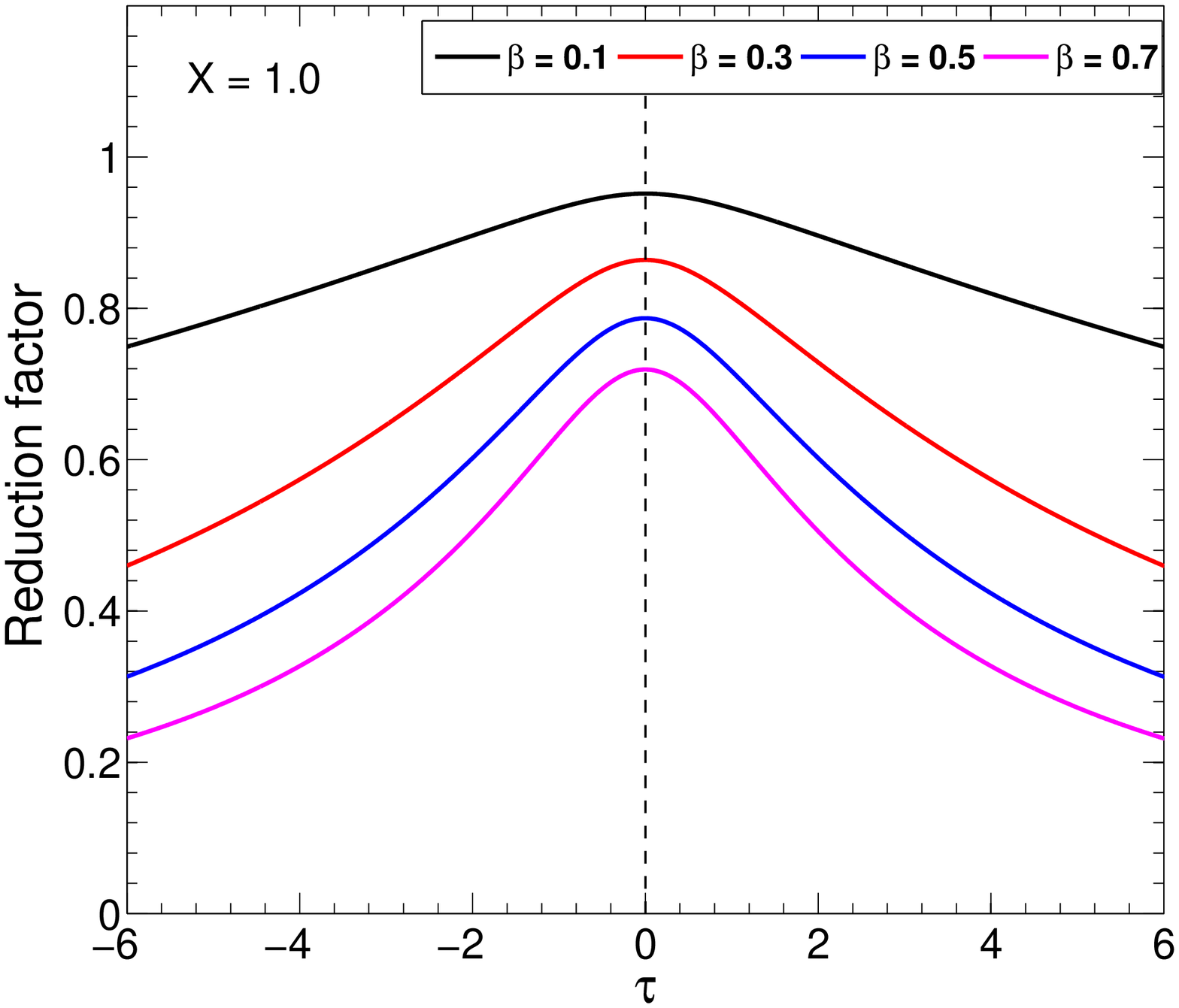}
        \vspace{0.2 cm}
	{\bf{(b)}}\\
    \end{minipage}
\begin{minipage}{.32\textwidth}
        \centering
        \includegraphics[height=2.1in,width=2.3in]{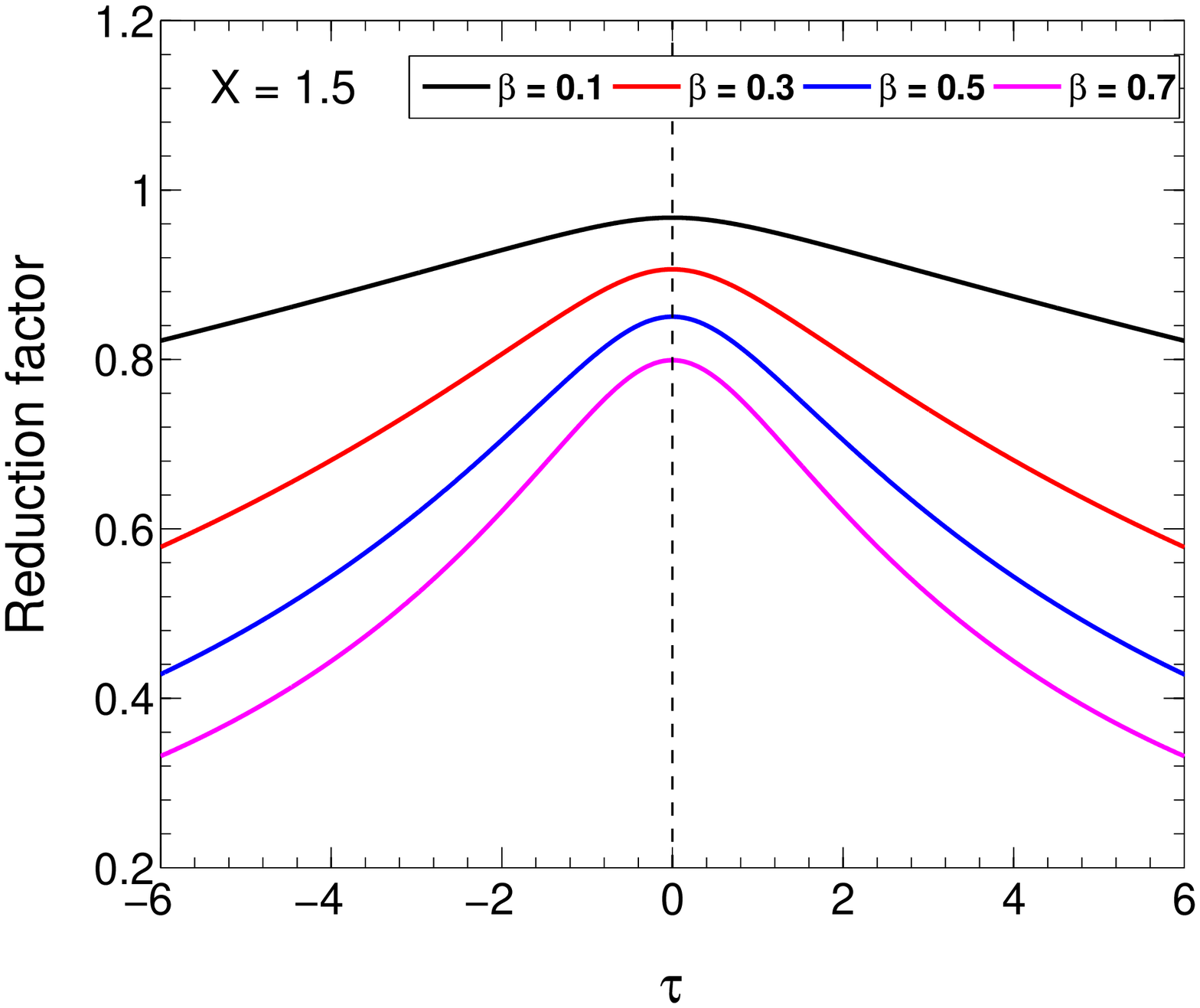}
        \vspace{0.2 cm}
	{\bf{(c)}}\\
    \end{minipage}
    \caption{Reduction factor, $\delta$ for surface density (see Eq. 8), plotted as function of $\tau$, the dimensionless measure of time in the sheared frame, for different finite thickness ($\beta$) and $X$ values. (a) shows the case for $X =0.5$, (b) for $X =1.0$, and (c) shows the case for $X =1.5$. As seen clearly, the reduction is important for high $|\tau|$ values, and for smaller values of $X$, for details see text.}
\label{fig:comparison_beta}
\end{figure*}

Next, we investigated how the reduction factor due to finite thickness depends on $\tau$ values. This is shown in Fig.~\ref{fig:comparison_beta}.

Due to the dependence of the reduction in self-gravity for finite thickness with $\tau$ values, at larger $\tau$ values (where the maximum amplification was likely to take place) the reduction is more for larger $\beta_{\rm s} (\ge 0.5)$ values (see Fig.~{\ref{fig:comparison_beta}} b), and hence the resulting self-gravity term fails to dominate over the pressure term, unlike the case for one-fluid system where the resulting self-gravity could still dominate over the pressure term; thus resulting in finite swing amplification even for $\beta_{\rm s} \ge 0.5$. Thus, the effect of gas on the swing amplification to happen at a later epoch (i.e. larger $\tau$ values) in turn indirectly limits the ranges of parameter for which the stellar disk will be able to support swing amplification.

To further study the mutual interplay between the effect of gas and of finite thickness, next we chose a higher value of gas-fraction ($\epsilon =0.15$) while keeping other input parameters as before, and calculated the resulting MAFs in both the stellar and gas disk from Equations  (\ref{swing2semi_one}) and (\ref{swing2semi_two}). We note that for this assumed set of input parameters, the inequality as given in Equation (\ref{cond-axisymmetric_semi}) is not satisfied, and consequently the swing amplification in the resulting two-fluid system will be high. The solution for $\beta_{\rm s} =0.5$ is shown in the Fig.~\ref{fig:swing2semi_compare} c.

We find that even for higher value of gas-fraction ($\epsilon$), the MAF of the swing-amplified features in the gas disk continues to decrease monotonically with the increase of finite thickness of the stellar disk. To express quantitatively, the MAF in the gas disk decreases by $\sim$ 65 \% for $\beta_{\rm s} =0.7$ when compared with the infinitesimally-thin disk. However, due to the larger contribution of gas, for a fixed $\beta$ value, the corresponding MAF values of the swing-amplified features in both the stellar and the gas disk are higher than that for $\epsilon = 0.1$. For larger $\tau$ values, the solution for stellar disk follows the solution for the gas disk, and oscillates around a non-zero mean; thus producing scalloped features \citep[for details see][]{Jog92}. Interestingly, due to the higher contribution from gas, the resulting two-fluid disk allows finite though small swing amplification in the stellar disk for $\beta_{\rm s} =0.5$ (see Fig.~\ref{fig:swing2semi_compare} c) unlike the case for $\epsilon =0.1$. This clearly brings out the mutual interplay between the opposite effects of finite thickness and the gravitational coupling of the fluids. Due to the larger contribution of gas, the resulting self-gravity term can dominate over the pressure term even for $\beta_{\rm s}=0.5$ (unlike the case of $\epsilon =0.1$); thus allowing finite amplification to take place. However, we find that for larger values of $\beta_{\rm s}(\ge 0.7)$ the resulting self-gravity term can no longer dominate over the pressure term and swing amplification in the stellar disk is prevented, that is, the effect of finite thickness prevails again over the effect of gravitational coupling of the fluids, but for a higher cut-off in the stellar thickness.

Thus, the dependence of reduction factor (due to finite thickness) on $\tau$ and the strong gravitational-coupling between two fluids, taken together can produce a wide range of complex yet rich physical scenarios which otherwise could not be captured by either the one-fluid or the two-fluid infinitesimally-thin modelling of the galactic disk.

\subsubsection{Both the stellar \& gas disk have finite thickness}

Here we studied a more realistic model for galactic disk where the disk is treated as a gravitationally-coupled two-fluid (stars and gas) system and each fluid has a total thickness of $2h_i$, $i $= s, g for stars and gas, respectively where $h_{\rm s} > h_{\rm g}$.

We considered a case where $Q_{\rm s} = 1.8$, $Q_{\rm g} = 1.2$, $\epsilon =0.15$, $\beta_{\rm g}=0.1$ (corresponding to thickness of 100 pc), $\eta= 1$ and $X=1$. This is representative of inner regions of Sb-type galaxies \citep{Jog92}. Then we varied the thickness of the stellar disk from $\beta_{\rm s} = 0.1$ to $\beta_{\rm s}=0.7$. This  allows us to investigate further the mutual interplay between the effect of finite thickness and the effect of gas for a more realistic model for galactic disk. Here we chose a slightly higher value for $Q_{\rm s}$ just to make sure that the resulting two-fluid system will satisfy the inequality given by Equation~(\ref{cond-axisymmetric}).

We note that, the scale-height of $HI$ disk in our Galaxy is $\sim$ 100-120 pc for $R< 8.5$ kpc \citep{Lock84}. Similarly, the scale-height of $H_2$ disk in our galaxy is $\sim$ 100 pc \citep[e.g. see][]{ScSa87,Wou90}. This supports our choice of $\beta_{\rm g}$ for the gas disk as reasonable.

\begin{figure}
    \includegraphics[width=0.85\linewidth]{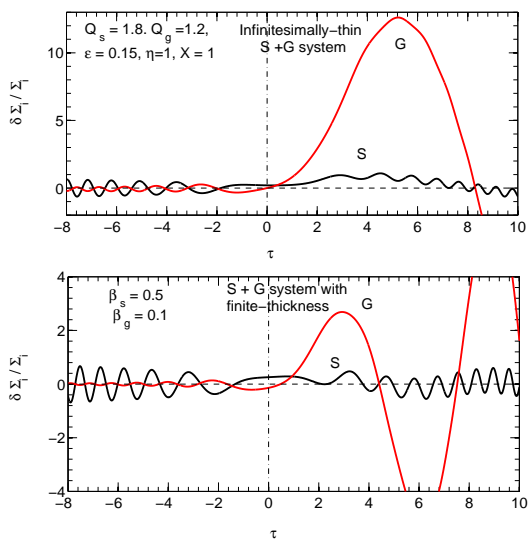}
\caption{Variation in $\theta =\delta\Sigma/\Sigma$, the ratio of the perturbation surface density to the unperturbed surface density, with $\tau$, dimensionless time in the sheared frame, plotted for gravitationally-coupled two-fluid (stars and gas) system. In the top panel, both fluids are modelled as {\it infinitesimally-thin} whereas in the bottom panel, both fluid disks have finite thickness. In both cases, $Q_{\rm s} = 1.8$, $Q_{\rm g}= 1.2$, $\epsilon = 0.15$, $\eta =1$, and $X =1$. For the infinitesimally-thin case (top panel), a high gas-fraction induces finite swing amplification in the stellar disk, but the introduction of finite thickness prevents the growth of swing-amplified modes almost completely (bottom panel). The net amplitude $\alpha \theta$ $\ll 1$at all $\tau$, where $\alpha$ is a scale factor.}
\label{fig:full_finite}
\end{figure}

The MAF in both the stellar and the gas disk continues to decrease with the introduction of finite thickness of both the disks, in agreement with the findings of previous sections. For illustration, the solution for $\beta_{\rm s} =0.5$ and $\beta_{\rm g} =0.1$ is shown in Fig.~\ref{fig:full_finite} and the solution for the infinitesimally-thin case is also plotted for the direct comparison \citep[also see][]{Jog92}.

Quantitatively, the MAF in gas disk is decreased by $\sim$ 60 \% for $\beta_{\rm g} =0.1$ and $\beta_{\rm s} = 0.7$ when compared with the two-fluid, infinitesimally-thin disk case. This is because of the reduction in the self-gravity of the joint system due to finite thickness of both stellar and gas disks. For the stellar disk, values of $\beta_{\rm s} \ge 0.3$ prevent the finite swing amplification almost completely, thus leaving the inner regions of the stellar disk devoid of any strong, small-scale spiral features. Even for some cases with high gas-fraction (e.g. $\epsilon =0.2$), the effect of the interstellar gas in supporting strong spiral arms in the stellar gas can not prevail over the effect of the finite thickness in suppressing the spiral arms in the stellar disk.

\subsection {Dependence of the result on $X$ }

So far we have used the normalized wavelength, $X=\lambda_y/\lambda_{\rm crit}=1$ to obtain the solutions for the swing amplification. Past literature has shown that the MAF of the swing-amplified features vary with different values of $X$ \citep[e.g see][]{Too81,Ath84}. Also, the reduction in self-gravity when written in terms of the dimensionless quantities, also depends on $X$ (see \S~2.2). Therefore, it is worth checking how the reduction in self-gravity due to finite thickness changes the resulting swing amplification for $X$ values other than unity.

In Fig.~\ref{fig:comparison_beta}, we have already shown how the reduction in the self-gravity changes for different $\beta$ values and for $X=0.5$ and $X=1.5$. We note that, the reduction in the self-gravity (at the mid-plane, $z=0$) depends on the argument $X^{-1} \beta$ (see \S~2.2), and therefore, for a fixed value of $\beta$, a smaller value of $X$ will cause a higher reduction in the self-gravity, and vice versa. This fact is evident when the reduction factors for $X=0.5$ and $X=1.5$ (plotted in Figs.~\ref{fig:comparison_beta} a \& c) are compared with the reduction factor for $X=1$ (see Fig.~\ref{fig:comparison_beta} b).

\begin{figure}
\centering
\includegraphics[width=0.85\linewidth]{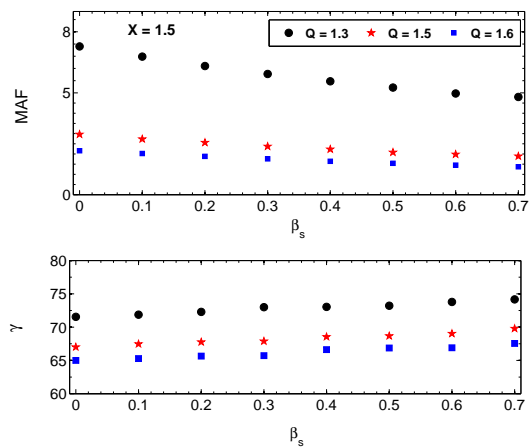}
\caption{Systematic variation of maximum amplification factor (MAF; for definition see \S~2.3.1) and the pitch angle ($\gamma$) of the resulting swing amplification, plotted in the top panel and bottom panel, respectively, as a function of disk finite thickness ($\beta$) for different Toomre $Q$ values, and for $X=1.5$, for a one-fluid disk. As seen clearly, the monotonic decrease in MAF due to the introduction of finite thickness also holds true for $X$ values other than $X=1$.}
\label{fig:study_x15}
\end{figure}

Now, we check the dependence of the results on $X$ for one-fluid model of the galactic disk.
 Fig.~\ref{fig:study_x15} shows the systematic variation in the MAF and the pitch angle ($\gamma$) of the resulting swing-amplification in a one-fluid system, for different finite thickness ($\beta$) values, and obtained for $X =1.5$. The MAF of the swing-amplified features decrease monotonically with the increase of finite thickness, and the pitch angle ($\gamma$) will remain mostly unchanged. Quantitatively, for $Q =1.3$, the MAF decreases by $\sim$ 34 \% for a thickness of 700  pc (i.e. $\beta =0.7$) when compared to the infinitesimally-thin disk. We checked that the relative decrease in MAF remains similar for other $Q$ values. This trend is in agreement with the findings for $X=1$. 
Also, a one-to-one comparison with Figs.~\ref{fig:fig2} and \ref{fig:study_x15} demonstrates that for larger $X$ values, the reduction in the self-gravity is smaller, and consequently the MAF is higher, as argued above in this section.

On the other hand, for $X=0.5$ and for $Q \ge 1.5$, the system could no longer support any swing amplification, because the reduction in self-gravity becomes large enough to swamp the finite swing amplification almost completely.

\section {Application to real galaxies}

In this section, we chose some parameter ranges ($Q_{\rm s}$, $Q_{\rm g}$, and $\epsilon$) typical of different regions of realistic galaxy cases and studied the effect of finite thickness on the resulting swing amplification process.

\begin{figure}
\centering
\includegraphics[height=2.5in, width=3.5in]{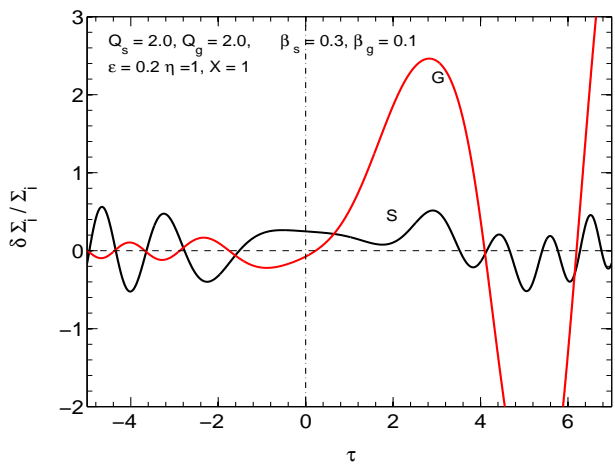}
\caption{Variation in $\theta =\delta\Sigma/\Sigma$, the ratio of the perturbation surface density to the unperturbed surface density, with $\tau$, dimensionless time in the sheared frame, plotted for a gravitationally-coupled two-fluid (stars and gas) system where both fluids have finite thickness. Here, $Q_{\rm s} = 2.0$, $Q_{\rm g} =2.0$, and $\epsilon =0.2$. The effect of finite thickness prevails over the effect of the interstellar gas, even for this case with high gas-fraction, and hence the stellar disk can no longer host strong, small-scale spiral arms (for details see text). The net amplitude $\alpha \theta$ $\ll 1$at all $\tau$, where $\alpha$ is a scale factor. }
\label{fig:case1_application}
\end{figure}

First, we chose $Q_{\rm s} =2.0$, $Q_{\rm g} = 2.0$, and $\epsilon =0.2$, $\eta=1$, $X=1$. This parameter range may be typical for the outer regions of the disks of Magellanic-type irregular galaxies where the gas fractions are high \citep[$\sim 20-30 \%$, e.g. see][]{GH84} and the lower values of $\kappa$ and $\Sigma$ will produce a larger value of Toomre $Q$ \citep[for details see][]{Jog92}. Then we set $\beta_{\rm g} =0.1$ and $\beta_{\rm s}=$ 0.3, and 0.5, and study the effect of the finite thickness on the resulting swing amplification in both the stellar and the gas disks. A typical case for $\beta_{\rm s} = 0.3$ and $\beta_{\rm g} = 0.1$ is shown in Fig.~\ref{fig:case1_application}, for illustrative purpose.

From Fig.~\ref{fig:case1_application}, it is evident that the stellar disk is not capable of showing any finite swing amplification, implying that in the outer parts of such Magellanic-type irregular galaxies, stellar disk can no longer display any strong, small-scale spiral features. This is a drastically different result when compared to the infinitesimally-thin two-fluid case in which the stellar disk can show finite swing amplification in the presence of high-fraction ($\epsilon \sim 20 \%$) of low velocity dispersion component, namely, the interstellar gas \citep[see Fig. ~2 in][]{Jog92}. The effect of finite thickness suppresses the swing amplification in the stellar disk, and even the high gas content can no longer support swing amplification in the stellar disk which otherwise would have been possible for the infinitesimally-thin system. Also, the MAF of the resulting swing amplification in the gas disk decreases as compared to what is found for the infinitesimally-thin system.

Also, in the cases of dwarf galaxies, the thickness of the gas disk is higher \citep{Ban11} and hence $\beta_{\rm g} =0.3$ is also possible. This will also lead to a further decrease in the amplification seen in the gas disk, and therefore may explain why such late-type dwarf galaxies do not show strong, local spiral features.

Next, we considered $Q_{\rm s} =1.8$, $Q_{\rm g} = 1.2$, and $\epsilon =0.15$. This parameter range is typical for the $R \sim$ 5 kpc of our Galaxy, where the molecular ring peaks, and perhaps also for the central regions of the gas-rich galaxies \citep[e.g. see][]{JS84,Jog92, BM98}. Then we set $\beta_{\rm s} =0.5$, $\beta_{\rm g} = 0.1$, and studied the effect of finite thickness on the swing amplification process. As already shown in Fig.~\ref{fig:full_finite}, even a high gas-fraction can not promote swing-amplified features in the stellar disk (contrary to the case for the infinitesimally-thin disk case, \citet{Jog92}). Here, the damping effect of finite thickness prevails over the effect of gas, and hence the stellar disk can no longer host strong, small-scale, swing-amplified spiral arms in the very central regions.

\section{Discussion}

Here, we discuss a few points relevant for this work.

1. The modelling of a galactic disk as a gravitationally-coupled two-fluid system where each component has a finite thickness (as done here) is quite general and can be applied to any two dynamically distinct populations of a galactic disk other than stars and gas system which we have explored here. The coexistence of two mutually opposite effects, namely, the effect of finite thickness in suppressing the swing amplification and the effect of a lower velocity dispersion component (e.g. gas) in promoting swing amplification can lead to diverse and complex physical results depending on the relative strengths of these two effects mentioned above.

2. Existence of a thick-disk component in disk galaxies was first discovered by \citet{Tsi79} and identified as a distinct structural component by \citet{Bur79}. After that several careful and extensive observations have revealed the ubiquity of such a thick-disk component for the external galaxies \citep[e. g. see][]{YD06,Com11a} as well as for Milky Way \citet{GR83}, consisting of relatively old and metal-poor stars  \citep[e. g. see][]{RM93,CB00}. Also, a recent study by \citet{Com17} has shown that the thick-disk component is not the artefact of the scatter diffuse light as has been suggested earlier in the literature. In this paper, we do not explicitly study the thick-disk case, however, we expect that the qualitative trends shown in this paper would hold good for the thick-disk case also.  

3. The stellar disk is known to flare (i.e. the scale-height increases) by a factor of few within the optical radius \citep[e.g.][]{NJ02b,deg97,Lop14}, and also the gas disk is known to flare steeply in the outer Galaxy \citep{Lev08}. Therefore, in the light of findings of this paper, the phenomena of flaring of both the stellar and the gas disks will have a suppressing effect on swing amplification operating in the outer parts of the optical disk and beyond. Thus we predict that the very outer parts of galaxies would not tend to support the small-scale spiral arms.

4. Some early-type dwarf galaxies in Virgo cluster (e.g. IC~3328) do show weak grand-design spiral arms despite having a thick disk and being gas-poor \citep{Jer00,Lis06,Lis09}. Therefore, in light of the findings from this paper, the presence of spiral arms in such systems is a puzzle. However, we point out that finite-thickness of the disk and dearth of interstellar gas does not necessarily rule out the possibility of an occasional spiral arm arising due to a tidal encounter \citep[e.g. as shown in ][]{Too72,BT87}. However, due to the reduction in self-gravity (because of finite thickness) in the disk mid-plane, the disk will be more stable, and hence the resulting spiral features will be in general weak. We also note that the structure seen in this galaxy (IC~3328) is global spiral pattern while we are considering only the local, small-scale spiral features.

5. Galaxies at high redshift are known be gas-rich \citep[some cases as high as $\sim$ 50 \%, see e.g.][]{Dad10,Tac10}, and also has thick disk \citep[e.g. see][]{Elm06}. Therefore, it would have been interesting to extend the analyses of this paper to those cases. However, we point out that these galaxies at high redshift display a wide variety of morphology from clumpy, disturbed disks to even chain-like structure \citep{Elm09}. Also, the turbulent velocities are very high and can be comparable to the underlying rotation velocity \citep{For06,Bour08}. Thus, these galaxies do not seem to have a well-defined differentially rotating disk as we encounter in galaxies in the local Universe. Since, differential rotation of the disk is a pre-requisite for the swing amplification to work, therefore it is not straightforward to extend the study presented in this paper for galaxies at high redshift.

6. For simplicity of the calculation, we have assumed a constant density along the $z$ direction throughout the calculation, and considered the finite thickness of the disk in such a way that thickness although finite, however is small when compared to the wavelength of the perturbation \citep[for more discussion see][]{Jog14}. This assumption, in turn restricts us from exploring the entire parametric space for $\beta$. Nevertheless, even the smaller $\beta$ range that we could explore here, has conclusively brought out the physical importance of the finite thickness on the swing amplification process.

7. We note that the formalism of introduction of finite thickness of the disk in the swing amplification process and the subsequent results obtained in the previous sections are essentially based on linear perturbation analysis. In reality, processes operating in the disk galaxies are non-linear in nature. In the past, it was shown that the presence of  non-linearity in the system can modify the results expected from the linear theory. For example, \citet{Don13} showed that the highly non-linear response of the galactic disk can significantly modify the persistence of the spiral structure. 

Here, we point out that recent $N$-body models of the galactic disk assume a finite thickness for the disk component. Therefore in the light of the finding of this paper, we caution that a careful choice is to be made for the thickness of the galactic disk in cases of simulations where the results predict the existence and the strength of the spiral structure.

\section{Conclusion}

In summary, we have studied the physical effect of finite thickness of a differentially rotating galactic disk on the resulting growth of non-axisymmetric perturbations via swing amplification. This was done by modelling the galactic disk first as a one-fluid system, and then as a gravitationally-coupled two-fluid (stars and gas) system. 

The main results of this work are summarized below.

\begin{itemize}

\item{The introduction of finite thickness decreases the maximum amplification factor (MAF) of the resulting swing-amplified spiral features  for the one-fluid model of the galactic disk.    
 This finding holds true for a wide range of Toomre $Q$ parameter values studied here.}
 
\item{The limiting value of Toomre $Q$ parameter denoting the sufficient condition for stability against non-axisymmetric perturbation is shown to get modified due to the introduction of the finite thickness of the galactic disk. The observed thickness range of 300-500 pc can suppress the growth of the non-axisymmetric perturbation at $Q \sim 1.7$ as compared to $Q =2.0$, required for the infinitesimally-thin disk case (as shown in the past literature).}

\item{For a gravitationally coupled two-fluid (star-gas) system, the net amplification is shown to be set by the mutual interplay between effect of gas in supporting the strong spiral features and the effect of finite thickness in preventing the spiral arms. We showed that even a high value of gas-fraction (e.g. $\epsilon = 0.2$) is not able to induce any finite swing amplification in the stellar disk with finite thickness. This is a drastically different scenario when compared to the two-fluid gas-rich infinitesimally-thin system.}

\end{itemize}

Spiral arms are known to transport angular momentum \citep{LYKA72,SJ14} and are one of the main drivers for the secular evolution of galaxies. Since the inclusion of finite thickness of disk can lead to the suppression of formation of strong, small-scale spiral arms (as shown here), therefore, finite thickness of a disk could have a non-trivial effect of delaying the long-term evolution of disk galaxies.

\bigskip
\noindent {\bf Acknowledgements:\\} 
We thank the anonymous referee for the useful suggestions which have helped to improve the paper.
SG acknowledges support from a Indo-French CEFIPRA project (PROJECT NO: `5804-1').
CJ would like to thank the DST, Government of India for support via a
J.C. Bose fellowship (SB/S2/JCB-31/2014).

\begin{appendix}
\section{Constraints on the range of parameter $\beta$}

In this work, the way the finite thickness of a galactic disk is included in the swing amplification process puts a constraint on the parameter range for $\beta$ that can be explored. The density distribution is taken to be constant along the $z$ direction. Therefore, although the disk has a finite thickness, the thickness has to be small as compared to the wavelength of the perturbation so that $kh \ll 1$, where $k$ is the wavenumber of the perturbation, and the disk has a total thickness of $2h$  \citep[for details see e.g. ][]{Jog14}. This in turn restricts us from exploring quantitatively the extreme cases where $kh$ exceeds unity.

Expressing the quantity $kh$ in terms of the dimensionless quantities introduced earlier in this paper, we get
$kh = X^{-1} \beta$. Therefore, we can explore the cases where the quantity $X^{-1} \beta$ does not exceed unity. Keeping this constraint in mind, we have adjusted the ranges for $\beta$ for different $X$ values considered here, so that the underlying assumptions taken to derive the equations in \S~2.2 are reasonably valid.

\end{appendix}

\end{document}